\documentstyle[preprint,prb,aps]{revtex}
\begin{document}
%\voffset=-1truein
\draft
\title{First order transition from antiferromagnetism to
ferromagnetism in Ce(Fe$_{0.96}$Al$_{0.04}$)$_2$}
\author{M. A. Manekar, S. Chaudhary, 
M. K. Chattopadhyay, K. J. Singh, S. B. Roy and P. Chaddah}
\address{Low Temperature Physics Laboratory,
Centre for Advanced Technology,\\ Indore 452013, India}
\date{\today}
\maketitle
\begin{abstract}
Results of dc magnetization study are presented  showing 
interesting thermomagnetic history effects across the       
antiferromagnetic to ferromagnetic transition in            
Ce(Fe$_{0.96}$Al$_{0.04}$)$_2$. Specifically, we observe 
(i)ZFC/FC irreversibility rising with increasing
field; (ii) virgin curve lying outside the envelope M-H curve.
We argue that these effects are quite different from the 
characteristics seen in spin-glasses or in
hard ferromagnets; they can be understood as metastabilities 
associated with a first order magnetic phase transition. 
\end{abstract}                          
\section{introduction}
The C15-Laves phase ferromagnetic compound CeFe$_2$, 
with its relatively low Curie temperature ( T$_C\approx$230K) and 
reduced magnetic moment ($\approx$ 2.3$\mu_B/f.u$)\cite{1}, 
is an unusual member of the RFe$_2$ (R= rare earth) family \cite{2}.
It is also known for quite some time that CeFe$_2$ is
on the verge of a magnetic instability \cite{3}. With 
small but suitable change in electronic structure on doping 
with elements like Co, Al, Ru, Ir, Os and Re at the Fe-site
of CeFe$_2$ (Ref.4), the higher temperature ferromagnetism  
readily gives in to a lower temperature 
antiferromagnetic phase, and after certain concentration of 
dopants (usually 5 to 10\%) the antiferromagnetic (AFM) 
phase replaces the ferromagnetic (FM) phase 
altogether\cite{5,6,7,8,9,10,11,12,13}. A recent neutron 
measurement has now confirmed the presence of antiferromagnetic 
fluctuations in the FM ordered state of pure CeFe$_2$ 
itself\cite{14}. This in turn suggests the existence of 
strong competition between the FM and AFM ground state in 
pure CeFe$_2$ with the AFM ground state being stabilized on 
doping. 

With the initial debate, whether the low temperature ground 
state of the doped CeFe$_2$ compounds is re-entrant          
spin-glass \cite{3,5,7} or antiferromagnet, being 
settled in favour of antiferromagnet \cite{8,10,11,12,13}, 
the more recent experimental efforts are mainly focussed on 
understanding the cause of this magnetic instability \cite{15,16}. The 
question being asked now whether the observed magnetic properties are 
linked to the instability of cerium electronic state and/or 
the peculiarity of the 3d-4f hybridization
\cite{15}, and a clear cut model explaining the interesting 
electromagnetic properties of CeFe$_2$ and its pseudobinary alloys
is yet to be established. There is one other aspect of 
the observed magnetic properties which  needs proper attention
and the information on which will be important for any future model, 
is the exact nature of the FM to AFM transition. Although it 
is generally believed that this transition is of first order 
in nature \cite{10,11,12}, no detail study exists in this 
regard. We have recently addressed this problem in Ru and Ir-doped 
CeFe$_2$ alloys \cite{17,18}.In this letter we shall 
focus on Al-doped CeFe$_2$ system. In contrast to
the quite sharp  FM-AFM transition in Ru, 
Co, Ir doped CeFe$_2$ alloys, this transition in Al-doped CeFe$_2$ alloys
is relatively gradual in nature \cite{8,11}. We report here 
interesting thermomagnetic properties of magnetization for 
a Ce(Fe$_{0.96}$Al$_{0.04}$)$_2$ alloy. 
These thermomagnetic properties are distinctly different from 
those observed in spin-glasses and  
hard ferromagnets. We argue that these thermomagnetic history 
effects in the present system
arise due to the first order nature of the FM-AFM transition. Such effects 
may be treated as characteristic signatures of a first 
order FM-AFM transition in general.
 
The sample used in the present study belongs to the same 
batch of samples used earlier in the study of bulk 
magnetic and transpport properties\cite{8}, 
and neutron measurements \cite{11}.The details of the   
preparation and characterization of the sample
can be found in Ref.8. We   
have used SQUID-magnetometer (Quantum Design-MPMS5) for    
measuring magnetization (M) as a  function of temperature 
(T) and applied magnetic field (H).We have 
checked the results varying scan length from 2 to 4 cm 
and no qualitative dependence on the scan length is found.
Before each experimental cycle the sample chamber was 
flushed with helium gas after heating it to 200K. This is 
to get rid of any residual oxygen leaking in the sample chamber over a 
period of time.

In Fig.1 we present M vs T curves, obtained      
both in the zero-field-cooled (ZFC) and field cooled (FC)   
mode, at various applied H. A sharp rise in M as a function of decreasing T 
indicates the transition from paramagnetic (PM) to FM 
state. This is followed by a sharp drop in M at a lower T 
indicating the onset of the AFM transition.
The PM to FM transition temperature 
(T$_C\approx$200K) and FM to AFM transition temperature 
(T$_N\approx$95K) obtained from the low field (20 Oe) M-T 
curve (see Fig. 1 (a)), agree well with those obtained 
earlier from ac-susceptibility measurements \cite{8}. With
the increase in H there is a marked decrease in T$_N$, 
and a relatively slow rise in T$_C$. A 
distinct thermomagnetic irreversibility (TMI)(i.e. 
M$_{ZFC}\neq M_{FC}$) is observed in the M-T curves with H=20 and 
100 Oe (see Fig. 1(a))  starting well inside the FM regime.
This kind of TMI is widely associated in literature with 
spin-glass transition\cite{19}, but can occur in a ferromagnet      
also if the measuring field H is of the order of the 
coercivity field \cite{20}. 
With the measured coercivity in the FM regime of our present sample 
being about 100 Oe, we attribute the 
observed TMI for H$\leq$ 100 Oe to the residual domain 
related pinning effects. 
In consonance with this conjecture, the TMI in the 
FM regime vanishes with the further increase in H (see Fig. 1(a) and (b)). 

While the TMI in the FM region vanishes for H$>$ 100 Oe, a 
distinct TMI emerges in the AFM region. 
In contrast with the TMI in the FM 
regime this increases in strength with the increasing H (see Fig. 1 (a) 
 and 1(b)). This is quite anomalous in comparison with the 
TMI associated with nonergodic behaviour of spin response in 
spin-glasses and hindrance of domain rotation and/or  
domain wall pinning in the ferromagnets. In both of these 
cases the TMI is known to reduce with the increase in the applied 
field. 

The anomalous field dependence of TMI above 1 kOe, we believe, is
not associated with pinning/hindered motions of the spins or magnetic domains,
but can be understood as due to metastabilities associated with 
FM-AFM transition being first order in nature \cite{21,22}. 
While cooling from the FM state to   
the AFM state, the FM state will continue to exist as 
supercooled metastable state below T$_N$ down to a certain  
metastability temperature T$^*$ \cite{21}. Between T$_N$ 
and T$^*$ fluctuations will help in the formation 
of droplets of the stable AFM state, and at T$^*$ 
an infinitisimal fluctuation will 
drive the whole system to the stable AFM state. 
While lowering T towards T$^*$ although the amount of 
metastable FM state will go on 
decreasing, the spin alignment and hence magnetization 
within the FM state will increase. These combined effects can give 
rise to nonmonotonic behaviour in the temperature dependence 
of M below T$_N$. This is quite evident in the higher  
field (H $>$ 10 kOe) M-T curves obtained in the FC mode (see Fig. 1(b)). 
In the ZFC mode the measurement  
always starts in the low temperature AFM state, and hence  
there is no contribution to M from the supercooled FM state. 
The implicit assumptions in the above arguments are: (a) 
the existence of a H dependent T$^*$(H) and (b) widening of the 
difference between T$_N$(H) and T$^*$(H) as a function of H.
It is worth mentioning here that a distinct hysteresis is observed 
also in the temperature dependence of resistivity 
of the same sample around T$_N$ measured in
the absence of any magnetic external field \cite{23}. 
Some support for FM-AFM phase coexistence across 
T$_N$ does exist from existing neutron studies on 
Ce(Fe,Al)$_2$ alloys \cite{11}. A detailed neutron 
measurement in presence of applied magnetic field will be 
very illuminating in this regard.

More support for the first order nature of the FM-AFM 
transition is obtained from the history effects we have 
observed in the isothermal field dependence of M.        
In Fig. 2 we plot M-H plots for Ce(Fe$_{0.96}$Al$_{0.04}$)$_2$
at various T. It is apparent from Fig.1 that FM order        
exists for T$\geq$100K and H$\geq$20 Oe, and the behaviour  
of M vs H at T=100K (see inset of Fig. 2(a)) is consistent 
with this picture. The technical saturation of M is reached 
quite early by H$\approx$3kOe  and the magnetization is 
quite reversible with coercivity field of $\approx$100 Oe. With 
lowering in T the nature of M-H curve changes drastically 
with the appearence of a hysteresis bubble. This hysteresis 
along with the observed cubic to rhombohedral transtion\cite{10,11},
have been associated earlier as possible signatures of       
the field induced first order metamagnetic transition from  
AFM to FM in Co-doped CeFe$_2$ alloys \cite{12}. We shall 
now elaborate more on this issue and argue that these 
hysteretic field dependence of magnetization is indeed      
associated with a first order phase transtion. 

Concentrating on the M-H curve at T=5K we find that
if the field excursion is confined to        
 H$_M=\pm$ 30 kOe, the M-H curve remains perfectly             
reversible. In this field regime the sample remains in the 
AFM state. The observed non-linearity in the low field 
($\leq$5 kOe) regime is due to a parasitic 
weak ferromagnetism \cite{24} leading to a canted spin 
state \cite{11}. When the applied H crosses the critical 
value H$_M$, M rises rapidly and upon reversal of H hysteresis is 
observed. H$_M$ is identified as the metamagnetic field for the 
onset of the FM order. The hysteresis loop, however, collapses 
 before H is reduced to zero, and reappears again on the 
third quadrant on reversal of H from beyond H$_M$ in the 
negative direction, giving rise to distinct  $\it{double}$  $\it{loop}$ 
structure. Qualitatively similar behaviour is observed at   
T=50K as well (see Fig. 2(b)). 
In ferroelectric  materials such $\it{double}$  $\it{loop}$ hysteresis 
is taken as the decisive evidence for a first order ferroelectric 
transition \cite{25}. We are unaware of similar emphatic        
arguments in the field of magnetic materials, 
although a field induced transition     
accompanied by hysteresis is often quoted to be a first   
order transition. 

While discussing the $\it{double}$  $\it{loop}$ hysteresis 
polarization curve in ferroelectrics, it was argued that 
such field induced first order transition can be explained in terms 
free energy curves obtained by expanding in a power series  
in polarization (P) and retaining only terms with even 
powers in P up to sixth order (Ref.25). The first inspection 
at the free energy curve thus obtained (see Fig. 8-16 of Ref.25), 
does not, however, provide a clear cut explanation of the 
hysteresis. This situation becomes more clear if we take 
recourse to a more modern treatment of such free energy 
curve (Fig. 4.6.1 of Ref.21). 
Phase coexistence  and metastability across 
the first order transition, and
hence hysteresis can be explained naturally from such a 
free energy curve. In our present magnetic material, on 
reduction of the field from field values above H$_M$, the high field 
FM state remains as a metastable state even in the low H 
regime, and this, we believe, is the cause of hysteresis. 
The residual ferromagnetic state remains    
even on reversal of the direction of H, leading to an       
 anomalous situation where the virgin curve lies outside 
the envelope curve.(The virgin curve in the reverse direction
is obtained after zero field cooling the sample from above  
T$_C$ followed by unidirectional increase of H in the 
negative direction). This anomalous behaviour of the 
envelope curve lying above the virgin curve remains  on 
reversal of H through zero in the positive direction. 
Virgin curve and the envelope curve overlap in the 
high H regime above H$_M$ (see Fig. 2(c) and inset    
of Fig. 2(b)). Such anomalous relation between the virgin and 
envelope curve is uncommon in magnetic materials, 
except for some recent reports in granular 
magnetic systems \cite{26}. However, in those materials 
the virgin curve goes outside the return envelope curve 
only after certain applied H and after that they do not seem to 
merge again. In these materials this anomalous behaviour is 
tentatively attributed to the prominent contribution from surface 
magnetism since such behaviour is observed only below       
a certain grain size\cite{26}. This is unlikely to be the case in 
our present bulk magnetic system, since the surface to volume ratio 
cannot be that high as in those granular magnetic materials.

In conclusion, the results of our present bulk
magnetization studies showing distinct thermomagnetic history 
effects, in conjunction with the existing 
information of structural distortion across the FM-AFM 
transition\cite{11}, strongly claim that the FM-AFM transition in the
Ce(Fe,Al)$_2$ system is of first order in nature. No thermomagnetic 
history effect is observed across the PM-FM                  
transition\cite{27} , and the various
features from the present as well as previous studies\cite{8,11}
indicate this transition to be a second order transition.
These information will be important in formulation of the 
theoretical model to explain the interesting magnetic 
properties of CeFe$_2$ and its pseudobinaries. We also 
present through our present study certain thermomagnetic 
properties which can be used to identify a first order 
ferromagnetic to antiferromagnetic transition in magnetic 
materials in general. We now propose 
a (H,T) path dependent (i.e. FC/ZFC mode) neutron 
measurement across the FM-AFM transition in Ce(Fe,Al)$_2$   
system, which should be  able to prove or disprove our 
claim of the existence of supercooled FM phase across T$_N$, hence the     
first order nature of the transition.

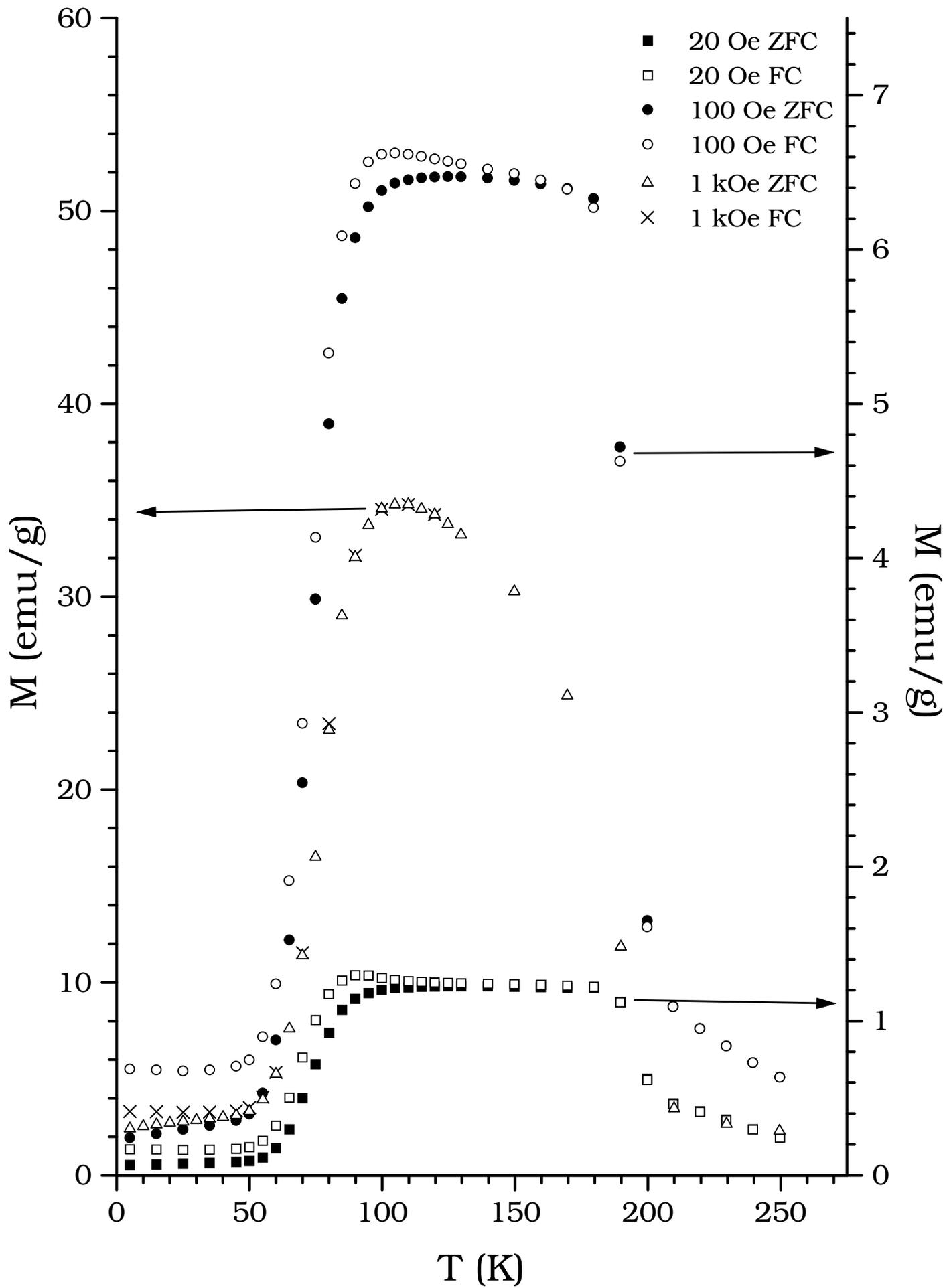
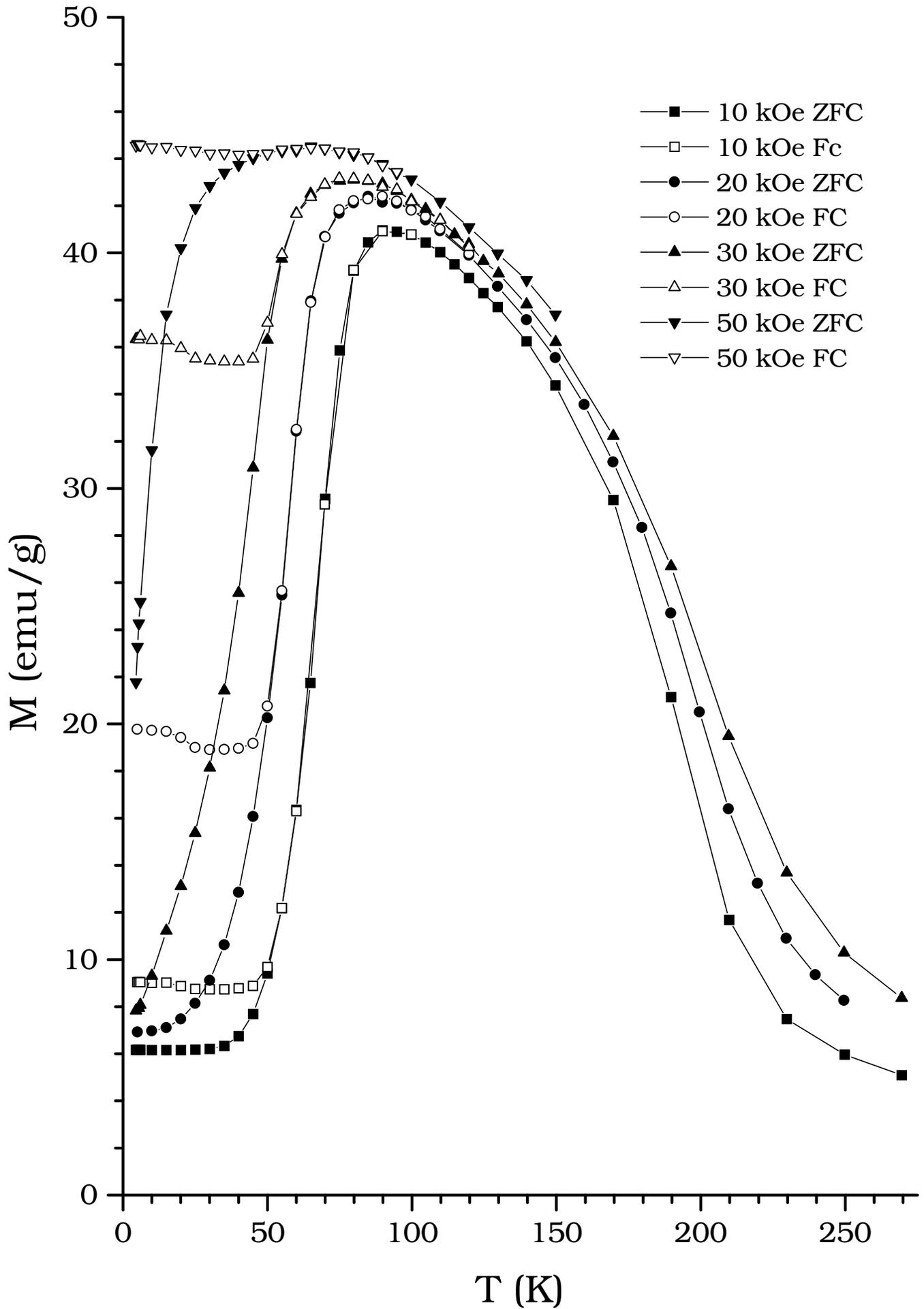
\begin{figure}
\caption{Magnetization vs temperature plots at various applied H, 
obtained both in the ZFC and FC mode.
Note that in the M-T curves with H=20 Oe and 100 Oe 
(Fig.1(a) the TMI extends well inside the FM regime.
In Fig. 1(b) lines     
serve as guide to the eyes.}
\end{figure}
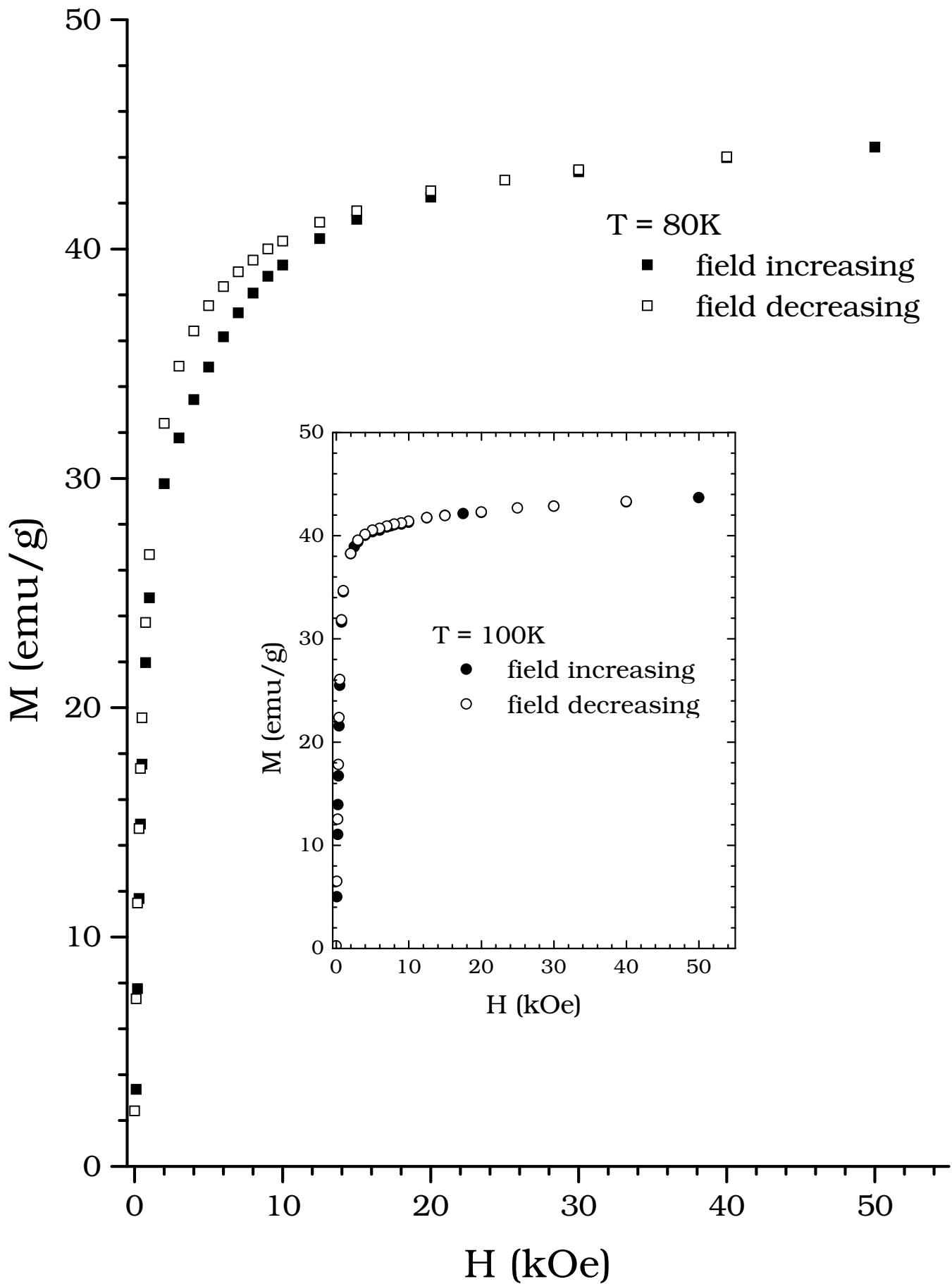
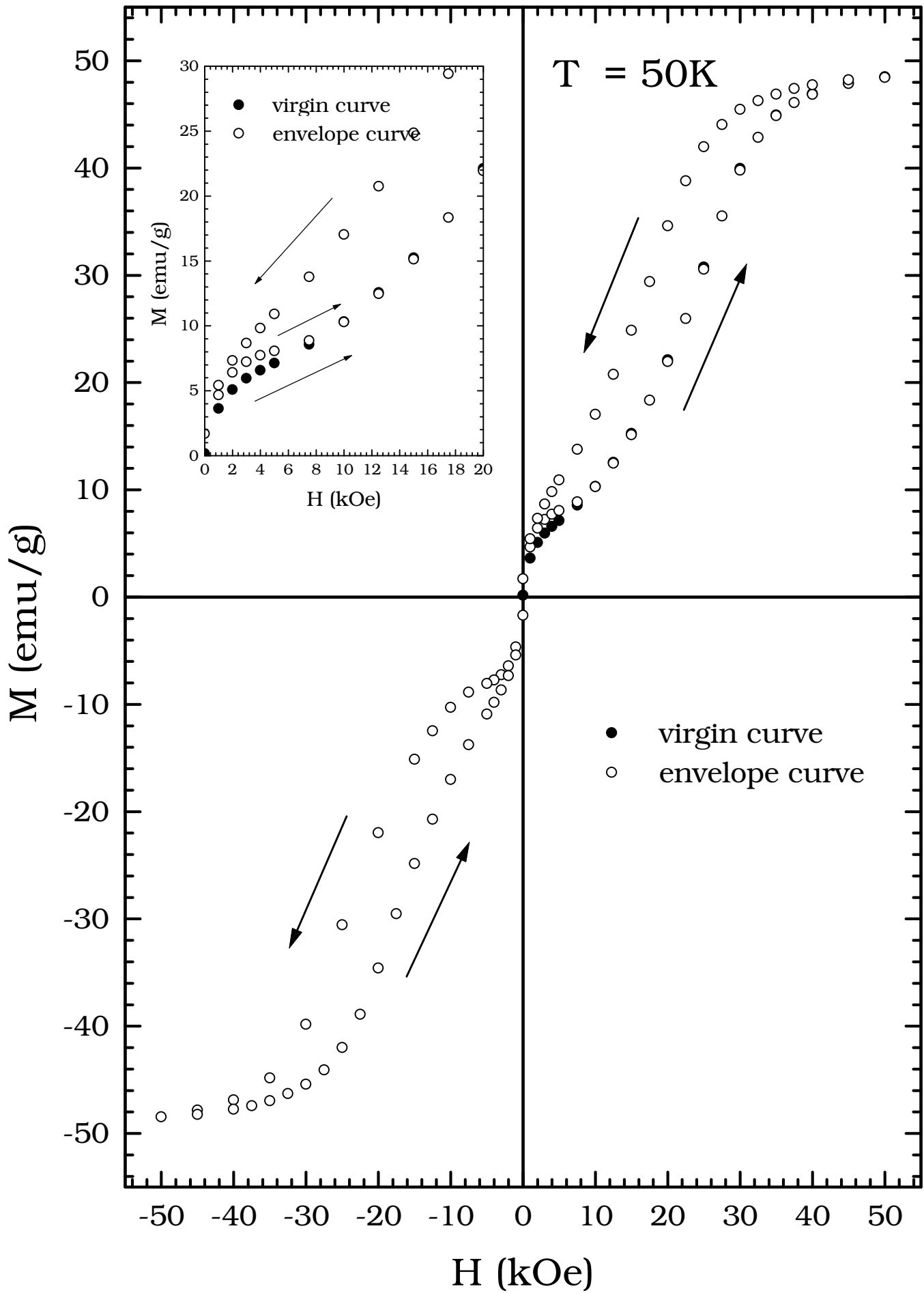
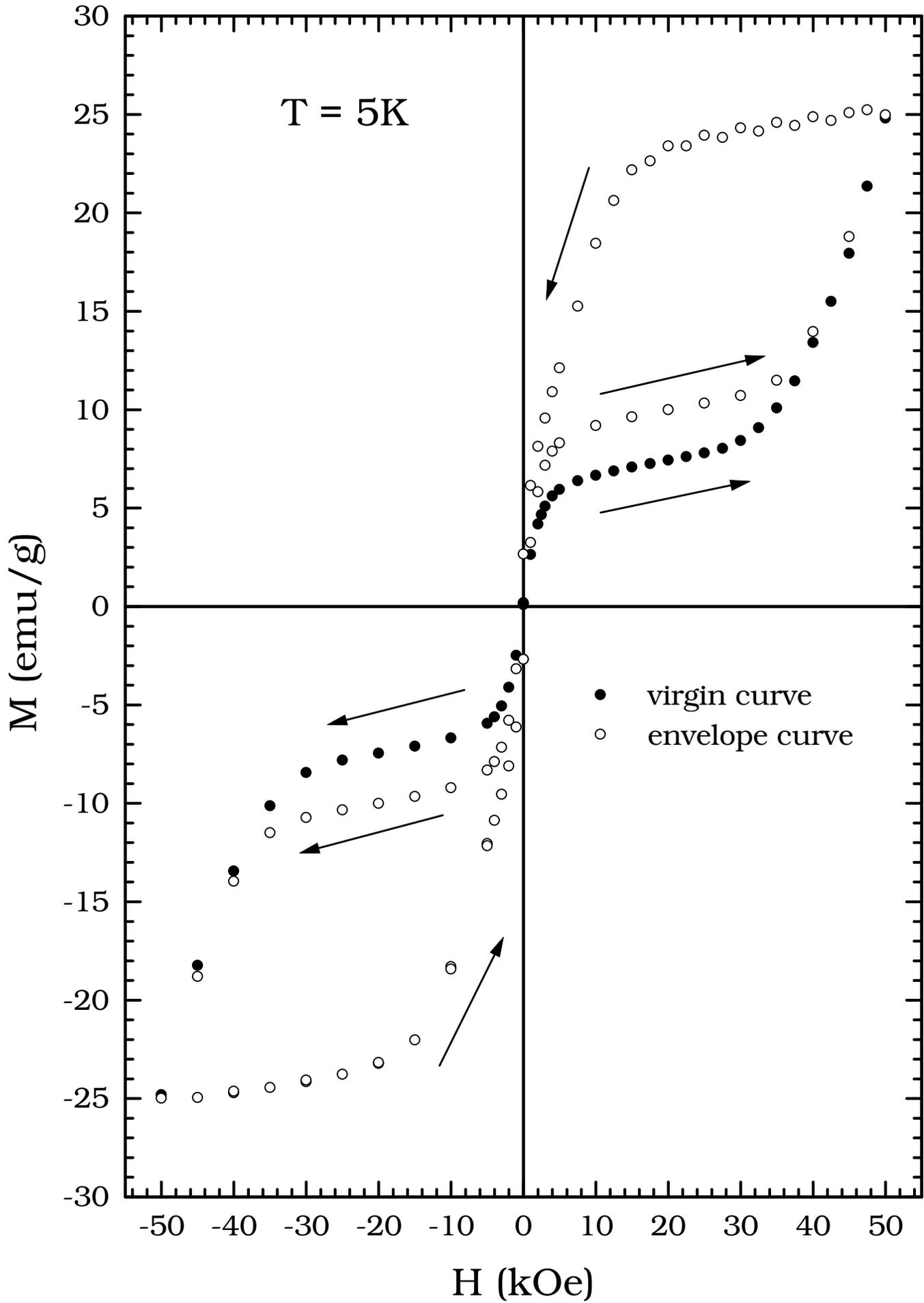
\begin{figure}
\caption{Magnetization vs field plots at various temperatures. Note that
in Fig. 2(b) (see the inset) and in Fig. 2(c) the virgin curve lies outside
the envelope hysteresis curve.}
\end{figure} 

\begin{references}
\bibitem{27}The TMI inside the ferromagnetic regime (see    
Fig. 1(a))  is observed only below a characteristic 
temperature T$_{irrv}$(H), which is distinctly separated 
from the PM-FM transition temperature.
\end{references}
\end{document}